# First principles investigation of the electronic structure of La$_2$MnNiO$_6$: A room temperature insulating ferromagnet

S. F. Matar[a], M.A. Subramanian[b], V. Eyert[c], M. Whangbo[d], A. Villesuzanne[a],

(a) Institut de Chimie de la Matière Condensée de Bordeaux, I.C.M.C.B-CNRS 87, Avenue du Docteur Albert Schweitzer, F-33608 Pessac Cedex, France.

(b) DuPont Central Research and Development, Experimental Station, E328/219, Wilmington, DE 19880-0328, USA.

(c) Institut für Physik, Universität Augsburg, 86135 Augsburg, Germany.

(d) Department of Chemistry, North Carolina State University, Raleigh, NC 27695-8204, USA.

***Abstract.***
*Using first principles calculations within DFT based on the full potential APW+lo method, we calculated the electronic and magnetic structures for the ferromagnetic and antiferromagnetic states of La$_2$MnNiO$_6$ and analyzed the site projected density of states and electronic band structures. Our calculations show that the ground state of La$_2$MnNiO$_6$ is ferromagnetic insulating with the magnetization in agreement with Hund's first rule and experimental findings.*

## 1. Introduction and theoretical framework

In the context of the emerging field of spin electronics, it is desirable to have semiconducting materials with strong ferromagnetic behavior near room temperature. However, the search for such materials is a difficult task due to conflicting requirements in the crystal structure, chemical bonding and electronic properties of semiconductors and ferromagnetic materials. CrO$_2$ is an example of halfmetallic ferromagnet at room temperature used in high density magnetic recording and has been extensively studied both experimentally and theoretically [1]. However, there are only a few examples of ferromagnetic insulators, which include the first discovered CrBr$_3$ [2], EuO, EuS and a Jahn-Teller system K$_2$CuF$_4$. This class of materials opens a wide realm for applications since the spins, electric charge and dielectric properties can be simultaneously tuned by magnetic and/or electric field. Recently, neutron diffraction studies confirmed a near room temperature ferromagnetism in La$_2$NiMnO$_6$ [3]. The purpose of this work is to present a theoretical analysis of the electronic and magnetic properties of $La_2NiMnO_6$ from first principles using a most accurate computational method within the framework of density functional theory (DFT), namely, the full potential APW + local orbitals method as implemented in the Wien*2k* package [4]. The validity and the limits of such a collective-electron treatment of strongly correlated oxide systems have been discussed by one of us in a recent review on magnetic oxides [5]. For the exchange-correlation potential, both the LDA parameterization of Perdew and Wang [6] and the GGA parametrization of Perdew, Burke and Ernzerhof [7] were used. Brillouin zone integrals were evaluated using the k-point mesh by Monkhorst and Pack with up to 234 irreducible points (11x11x7) [8]. The difference in total energies was converged to below 0.002 eV with respect to both k-point integration and kinetic energy cut-off, and to $\Delta E = 10^{-4}$ eV convergence between successive iterations.

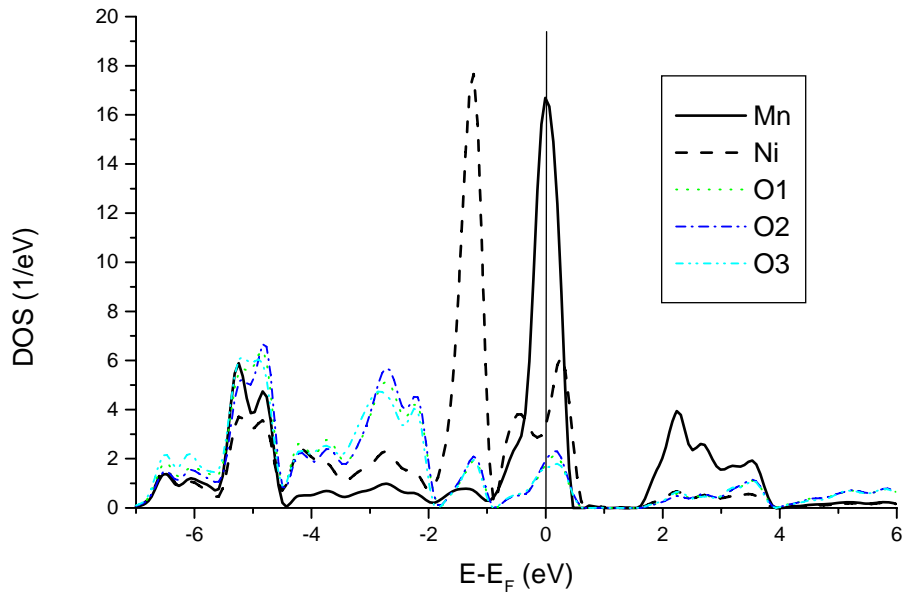

*Fig. 1: Site projected density of states in the non magnetic state of $La_2MnNiO_6$. Empty La(4f) states above $E_F$ are not shown for the sake of clarity.*

2. Results and discussion.

A. Search of magnetic instability: Non spin-polarized calculations.
In order to establish a reference for the spin-polarized calculations we started with a set of spin-degenerate calculations. While this does not represent a paramagnetic situation, which would require heavy computations with large supercells and random spin orientations, our objective here is to point out the possible magnetic instability on the different chemical species. The site projected DOS are shown in fig. 1. The valence band is dominated by oxygen states in the range of -7 to -2 eV; they mix with the lower part of Mn and Ni states. The Ni DOS are mainly below the Fermi level ($E_F$) while Mn states are found at $E_F$ ($t_{2g}$ states) and above $E_F$ (empty $e_g$ states). This picture is consistent with the expected oxidation states of $Ni^{2+}$ ($d^8$) and $Mn^{4+}$ ($d^3$). As a consequence, the DOS at $E_F$, $n(E_F)$, is significantly larger for Mn than for Ni. Within the Stoner theory of band ferromagnetism, which is a mean field approach, the large DOS at the Fermi level is related to the instability of the non magnetic state with respect to the onset of intraband spin polarization, when $n(E_F).I >1$. Here, I, is the Stoner integral, which was calculated and tabulated by Janak [9] for the elemental systems: I(Mn)=0.408 eV and I(Ni)=0.503 eV. With $n_{Mn}(E_F)$= 8.5 eV$^{-1}$ and $n_{Ni}(E_F)$= 2 eV$^{-1}$, the $n(E_F).I$ values are 3.47 and 1.06 for Mn and Ni, respectively. This suggests that both Mn and Ni 3d states are susceptible to spin polarization, as confirmed in our subsequent spin-polarized calculations (see below). From the large oxygen contribution at the Fermi energy, which results from strong metal 3d - oxygen 2p interactions, we expect the O 2p states to experience a finite spin-polarization as well.

B. Spin polarized calculations of $La_2NiMnO_6$.

In the spin-polarized calculations two different magnetic configurations were considered for the search of the ground state: (i) ferromagnetic ordering and (ii) a C-type antiferromagnetic ordering (i.e., ab-planes containing antiferromagnetically ordered transition metal cations are coupled ferromagnetically along the *c* direction). The results of these calculations using both the LSDA and the GGA are summarized in Table 1.

|  | LDA | GGA | Experimental |
|---|---|---|---|
| M (Mn/Ni) $\mu_B$ | F: 2.62 / 1.41<br><br>AF: ±2.41 /±1.37 | F: 2.64 / 1.46<br>AF: ±2.42 /±1.45 | F :  Mn : 3<br>Ni: 2 |
| M (O/La) $\mu_B$ | F: 0.1 / ~0<br>AF: 0 | F: 0.1 / ~0<br>AF: 0 | - |
| M (Total/fu) $\mu_B$ | F: 5<br>AF: 0 | F: 5<br>AF: 0 | ~5 |
| $\Delta$E (F-AF) eV/fu | -0.15 | -1.3 |  |

Table 1: FP-LAPW calculations for the ferromagnetic (F) and antiferromagnetic (AF) states of La$_2$MnNiO$_6$.

Although the atomic magnetic moments of Mn and Ni do not reach the experimental values, it is interesting to observe that the total magnetization per formula unit is 5$\mu_B$ in agreement with experimental findings. This is due to the fact that there is a small amount of spin polarization on the oxygen atoms, which sums up with the atomic moments of Ni and Mn to the integer value of 5. This value agrees with the Hund's first rule moments of Mn$^{4+}$ (3d$^3$): t$_{2g}^3$ e$_g^0$ and Ni$^{2+}$ (3d$^8$): t$_{2g}^6$ e$_g^2$. According to the Goodenough-Kanamori rule, the coupling between adjacent Mn$^{4+}$ and Ni$^{2+}$ sites should be ferromagnetic [10]. This prediction is confirmed here, since both functional calculations show that the ferromagnetic state is more stable than the C-type antiferromagnetic state in agreement with experiment [3]. Note that the energy difference obtained within GGA is quite large (1.3 eV per formula unit), so that a G-type antiferromagnetic ordering, which involves 50% more AF nearest-neighbor interactions than does a C-type antiferromagnetic ordering, is expected to be largely unstable. This point is under investigation. In the following we examine the electronic band structures in some detail.

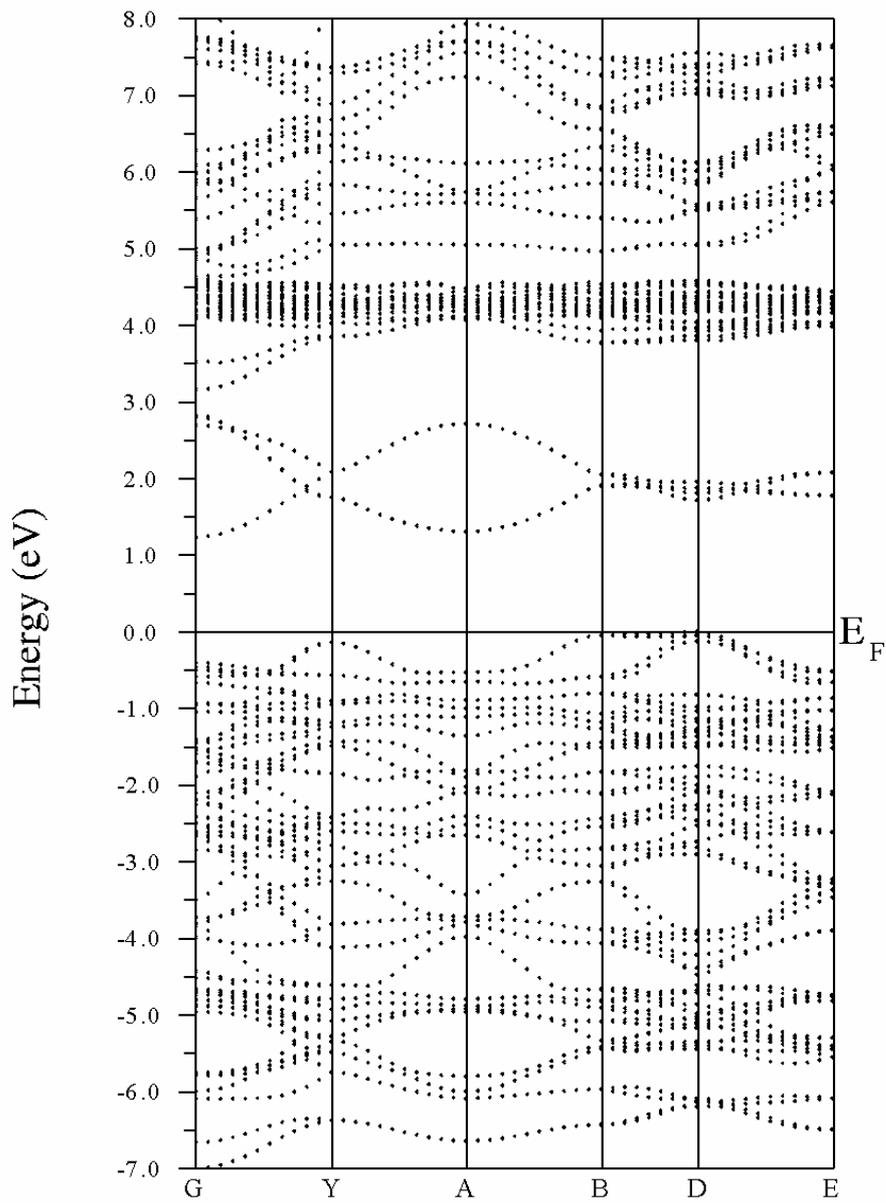

*2a*

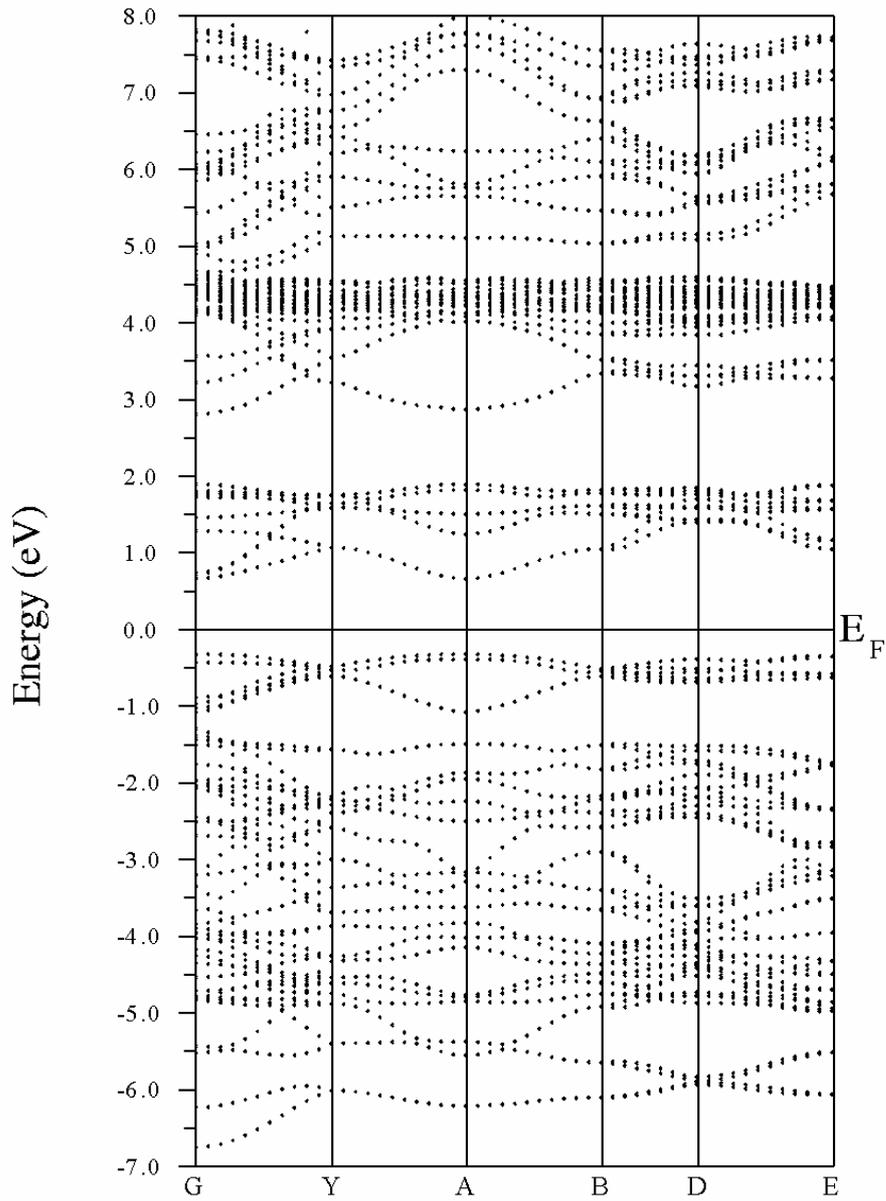

2b

*Figure 2: Band structure for the ferromagnetic state of $La_2MnNiO_6$. a) Spin ↑ bands; b) Spin ↓ bands along the main lines of the monoclinic BZ (G represents the $\Gamma$ point of the BZ).*

An important feature of the ferromagnetic state is the presence of a band gap of ~1 eV and an insulating behavior for both ↑ (majority spins) and ↓ (minority spins) bands (Figure 2). Among the empty conduction band, the La (4f) bands are flat and are found in the range 4 to 5 eV. The empty Mn ($e_g$↑) bands are located between these bands and $E_F$. All of the Mn ($t_{2g}$↑) bands lie below $E_F$ and are responsible for the 3 $\mu B$ moment, whereas the Ni ($e_g$↑) as well as the Ni ($t_{2g}$↑) bands are filled and lie below $E_F$. The panel of the minority spin bands is characterized by the empty Mn ($t_{2g}$↓), Mn ($e_g$↓) and Ni ($e_g$↓) bands between $E_F$ and the La (4f) bands. The Ni ($t_{2g}$↓) bands occur within the valence bands. These features, which are illustrated in the Mn and Ni DOS plots of figs. 3 a,b are fully consistent with the electronic structures expected from the oxidation of $Ni^{2+}$ and $Mn^{4+}$. The above analysis clearly shows different bonding behaviors of the two transition metal atoms in the majority and minority spin subbands. Such a feature of spin-dependent chemical bonding has been discussed before [5, 11, 12], and will be detailed further for this system in a full paper.

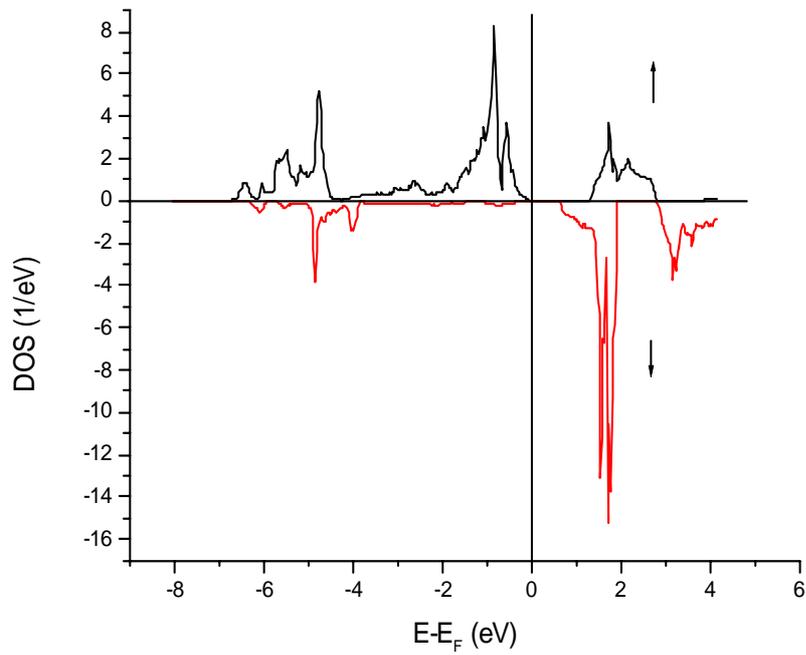

a)Mn

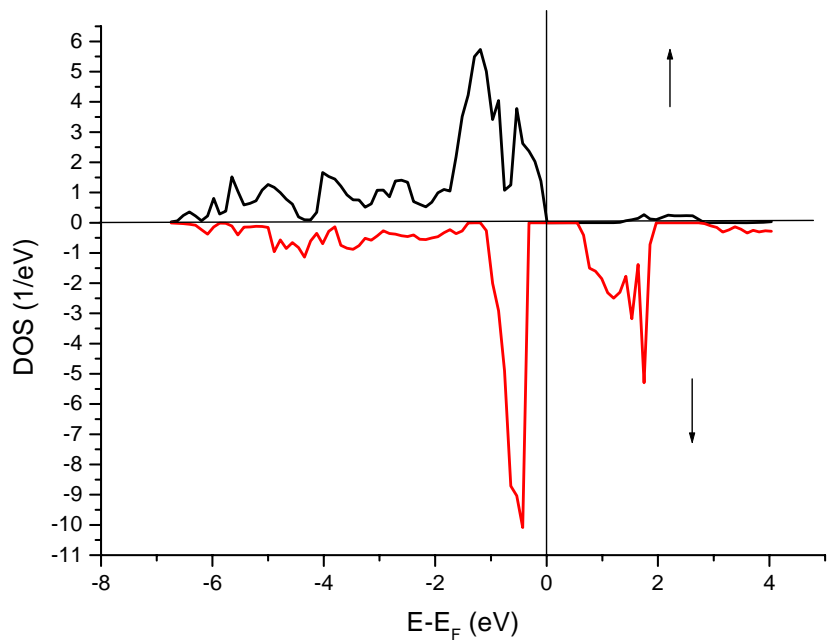

b)Ni

*Figure 3: Site and spin projected DOS of Mn and Ni within La$_2$MnNiO$_6$ system.*

3. Concluding remarks.

The electronic and magnetic properties of La$_2$MnNiO$_6$ were investigated using first principles DFT calculations. Spin-degenerate calculations clearly pointed to a magnetic instability involving both the Mn and Ni states. Indeed, our spin-polarized calculations show that the ground state of La$_2$MnNiO$_6$ is ferromagnetic with a magnetic moment of 5 $\mu_B$ per formula unit and is insulating with a band gap of 1 eV, in agreement with the available experimental data. A more detailed investigation on the relationship between the electronic structure, chemical bonding and the exchange coupling is underway.

4. Acknowledgements.

Computational facilities of the intensive scientific pole "M3PEC" of the University Bordeaux 1 using the supercomputer "Regatta" partly financed by the "Conseil Régional d'Aquitaine" are gratefully acknowledged. V.E. acknowledges support from the Deutsche Forschungsgemeinschaft through SFB 484, and M.-H. W thanks the financial support from the Office of Basic Energy Sciences, Division of Materials Sciences, U. S. Department of Energy, under Grant DE-FG02-86ER45259.